\documentclass[showpacs,preprintnumbers,amsmath,amssymb]{revtex4}
\usepackage{graphicx}

\begin{document}
\preprint{  CERN-PH-TH/2005-178}
\title{Can the cosmological constant  undergo abrupt changes?\vspace{0.2cm} }

\author{  Alejandro Cabo$^{1,2}$, Alejandro Garcia-Chung$^{3}$ and Alejandro Rosabal$^{4}$\vspace{0.2cm} }

\affiliation{$^{1}$ Theory Division, CERN, Geneva, Switzerland}

\affiliation{$^{2}$  Grupo de F\'{i}sica Te\'orica, Instituto de
Cibern\'etica, Matem\'atica y F\'{i}sica, Calle E, No. 309,
Vedado, La Habana, Cuba}

\affiliation{$^{3}$ Departamento de F\'{i}sica, Facultad de
Ciencias Naturales, Universidad de Oriente, Santiago de Cuba, Cuba
}

\affiliation{$^{4}$ Departamento de F\'{i}sica, Universidad de
Granma, Bayamo, Cuba. }

\begin{abstract}
\noindent The existence of a simple spherically symmetric and
static solution of the Einstein equations in the presence of a
cosmological constant vanishing outside a definite value of the
radial distance is investigated. A particular succession of field
configurations, which are solutions of the Einstein equations in
the presence of the considered cosmological term and auxiliary
external sources,  is constructed. Then, it is shown that the
associated succession of external sources tend to zero in the
sense of the generalized functions. The type of weak solution that
is found  becomes the deSitter homogeneous space-time for the
interior region,  and the Schwartzschild space in the outside
zone.

\ \bigskip
\end{abstract}


\maketitle

\section{\protect Introduction}

The gravitational equations in the presence of a cosmological
constant $ \Lambda$ have been the subject of continued attention
in the literature about\ Classical and Quantum Gravity,  \ since
their introduction by Einstein as a means to \ define a static and
homogeneous model for the Universe \cite {Einstein}. \  In
particular the question about the physical relevance of these
equations \cite{dymnikova, weinberg,douglass,peebles} have
received a large amount of interest in recent times, thanks to the
modern experimental evidence signalling the relevance of this
quantity in describing the expansion rate of the Universe
\cite{reiss, perlm}.

It is an accepted fact that the satisfaction of the Einstein
equations for spherically symmetric and static solutions forces
the value of the cosmological constant $\Lambda $ to be rigorously
constant, that is, independent of the radial coordinate. \ The
known solution in this situation is the deSitter space-time.
Inside this space, the matter is pushed by the gravitational force
away from the centre, much as it could be attracted to the origin,
in the interior of a Schwartzschild black-hole
\cite{weinbergbook}.

\ In this work we  discuss the possibility that the same original
Einstein equations in the presence of a cosmological term, \ could
show a form for it that would be  rigorously non-vanishing and
space-independent inside a certain sphere, but reducing to zero
outside it. This field configuration will then define the deSitter
solution as the internal space and the Schwartzschild solution as
the external one.

The main circumstance suggesting the existence of this solution is
the fact that, at the special radial distance in which the metric
becomes singular for both the deSitter and Schwartzschild
solutions, the non-linearity of the equations could allow for
solutions in the sense of the generalized functions, showing a
sudden change in $\Lambda. $ \ \

\ The plan of the work will be as follows. In Section 2, the
Einstein equations for a diagonal energy-momentum tensor are
written and the notation to be employed is defined. \ Section 3
continues by considering a  rough motivation for the existence of
the solution  after solving the basic radial and temporal \
Einstein equations. A main point is that, assuming the exact
equality between \ $-g_{00}$ and $1/g_{rr}$ the equations for
these quantities \ become linear  and are exactly satisfied in the
sense of the generalized functions by the piecewise defined
solution given by the Schwartzschild and the deSitter fields for
the external and internal regions respectively. As is known, for
this class of centrally symmetric problems, these two equations
are
sufficient to fully determine the only two unknown fields $g_{00}$ and $%
1/g_{rr}.$ Therefore, if the solution for these quantities would
result in being  to be non-vanishing and differentiable, the full
set of equations could be  automatically  solved. \ However, the \
non-linearity of the resting Einstein equation makes it  necessary
to check that the full set of Einstein equations can be considered
as solved in some generalized sense. \ For this purpose, in
Section 4,  a physically grounded  solubility criterion is defined
for the satisfaction of the  Einstein equations. It rests in the
natural assumption of considering as a weak solution of the
non-linear equations, the limit of  a particular succession of
field configurations, for which each element solve the equation in
the presence of external sources and for which, moreover, the
corresponding succession of sources tends to zero in the sense of
the generalized functions. The need for involving  a particular
succession in the definition comes from the assumed non-linearity
of the equations. In the case where the equations are linear, the
generalized functions, being the limit of the linear functionals
associated to each field configuration in the succession,  could
be considered as the solution. However, the non-linearity  makes
it necessary to have a more precise definition by including the
particular succession that allows the limit of the external
sources to be vanishing.

Finally in Section 5 it is shown that the defined solubility
criterion can be satisfied for the Einstein equations in the
presence of a cosmological constant which reduces to zero outside
the radial distance $r_{0}$ for which the deSitter temporal metric
component $g_{00}$ vanishes. Roughly speaking, we found a
particular succession of field configurations  that produce a
vanishing limit for the externals sources associated to each of
its elements, in the sense of the generalized functions.

It becomes clear that the resulting gravitational field
configuration shows a kind of naked singularity, since for
example, the radial derivative of the $g_{00}$ metric components
has a discontinuity at the boundary \cite{giambo,joshi,harada}.

\

The results are reviewed and commented in the conclusions.

\ \

\section{Einstein equations }

The squared line element for spherically symmetric systems will be
written in the form
\begin{eqnarray}
ds^{2} &=&\mathit{g}_{00}(r){dx^{0}}^{2}+g_{rr}\text{ }dr^{2}+g_{\phi \phi }%
\text{ }d\phi ^{2}+g_{\theta \theta }\text{ }d\theta ^{2},  \label{metric} \\
&=&-\mathit{v}(r){dx^{0}}^{2}+u(r)^{-1}dr^{2}+r^{2}(\text{sin}^{2}\theta \text{ }%
d\phi ^{2}+d\theta ^{2}),  \nonumber
\end{eqnarray}
in which the functions $u$ \ and $v$ \ are defined in terms of the \ metric
components as
\begin{eqnarray}
v(r) &=&-g_{00}=\exp (\nu (r)),  \label{nula} \\
u(r) &=&g_{rr}=\exp (-\lambda (r)).  \nonumber
\end{eqnarray}
We will consider for the start, a set of equations slightly
generalizing the usual Einstein equations with \ a cosmological
term. \ For the case of a static and spherically symmetrical
solution, when there are no external sources
$J_{0}^{0}$,$J_{r}^{r}$,$J_{\phi }^{\phi }$ and $J_{\theta
}^{\theta }$ acting on the system, the equations to be examined
cab be written in the form
\begin{widetext}
\begin{eqnarray}
J_{0}^{0} &=&-\Lambda _{0}(r)-\frac{{u^{\prime }(}r{)}}{r}+\frac{1-u(r)}{%
r^{2}}=-\Lambda _{0}(r)+G_{0}^{0}(r)=0,  \label{E1} \\
J_{r}^{r} &=&-\Lambda _{r}(r)-\frac{u(r)}{v(r)}\,\frac{{v^{\prime }(r)}}{r}+%
\frac{1-u(r)}{r^{2}}=-\Lambda _{r}(r)+G_{r}^{r}(r)=0,  \label{E2} \\
J_{\phi }^{\phi } &=&J_{\theta }^{\theta }=-\Lambda (r)-\frac{u(r)}{{2}}(%
\frac{v^{^{\prime \prime }}(r)}{v(r)}-\frac{v^{\prime }(r)^{2}}{2v(r)^{2}}+
\label{E3} \\
&&\frac{v^{\prime }(r)\text{ }u^{\prime }(r)}{2\text{ }v(r)\text{ }u(r)}+%
\frac{{1}}{r}(\frac{u^{\prime }(r)}{u(r)}+\frac{v^{\prime }(r)}{v(r)})),
\nonumber \\
&=&-\Lambda (r)+G_{\phi }^{\phi }(r),  \nonumber \\
\text{ \ \ } &=&-\Lambda (r)+G_{\theta }^{\theta }(r)=0,  \nonumber
\end{eqnarray}
\end{widetext} in which the Einstein tensor $G_{\mu }^{\nu }$ is
diagonal in the spherical coordinates, and \ its diagonal
components are given as
\begin{eqnarray*}
G_{0}^{0} &=&\Lambda _{0}(r), \\
G_{r}^{r} &=&\Lambda _{r}(r), \\
G_{\phi }^{\phi } &=&G_{\theta }^{\theta }=\Lambda (r){.}
\end{eqnarray*}
 These metric components are written in terms of the three functions
 $\Lambda _{0},\Lambda _{r}$ and
$\Lambda $, which only depend on the radial coordinate. \ Since
the vanishing of the covariant divergence of the Einstein tensor
is an identity for any field configuration (see\cite{synge}) it
follows that
\begin{widetext}
\[
G_{\mu \text{ };\text{ }\nu }^{\nu }(r)\text{\ \ }=\frac{1}{\sqrt{-g(r)}}%
\partial _{\nu }(\sqrt{-g(r)}G_{\mu \text{ }}^{\nu }(r))-\frac{1}{2}\partial
_{\mu }(g_{\gamma \nu }(r))\text{ }G_{\text{ }}^{\gamma \nu }(r)\text{ }=0,
\]
\end{widetext} As usual, $\ g$ is \ the $\det $er$\min $ant of the
metric tensor
\[
g(r)=-\frac{v(r)\text{ }r^{4}\text{sin}^{2}\theta \text{ }}{u(r)}.
\]

\section{Indications for the solution}

Let us motivate the existence of a solution of the Einstein
equation in the presence of a cosmological term that  does not
reduce itself to a fixed constant times the metric tensor for all
the points of the space-time. For this purpose it can be first
noticed that when all the $\Lambda _{0},\Lambda _{r}$ and $\Lambda
$ functions are selected as equal among them and to a given
constant $\Lambda $, the set of equations (\ref{E1})--(\ref{E3})
have the usual deSitter solution showing the regular behaviour at
the origin
\begin{eqnarray}
u(r) &=&1-\frac{\Lambda \,\text{\ }r^{2}}{3},  \label{zero} \\
&=&1-\frac{\phi _{0}^{2}\,\text{\ }r^{2}}{6},  \nonumber \\
\mathit{v}(r) &=&1-\frac{\Lambda \,\text{\ }r^{2}}{3} \\
&=&1-\frac{{{\mathit{\phi _{0}}^{2}}}\,r^{2}}{6},  \nonumber
\end{eqnarray}
in which the scalar field parameter $\phi _{0}\,$\ is related to
$\Lambda $ through
\[
\Lambda =\frac{\phi _{0}^{2}}{2},
\]
where $\phi _{0}^{2}$ can be interpreted as the square of the mass
times the square of the scalar field producing the same
cosmological term as the Klein-Gordon Lagrangian, when the field
is assumed to be a  constant in all the space. The reason for
introducing this parameter is the fact that the field
configuration discussed here has a close relationship with a
particular solution for the Einstein Klein Gordon system discussed
in  \cite{caboayon}.

Consider now the external region to the sphere having  radius
${\,}r_{0}$  and assume  that all three functions $\Lambda _{0},
\Lambda _{r}$ and $\Lambda $ in (\ref{zero}) vanish. Within this
region, \ the Schwartzschild solution
\begin{eqnarray}
u(r) &=&1-\frac{{\,}r_{0}}{r}{,} \\
v(r) &=&1-\frac{{\,}r_{0}}{r}{,}
\end{eqnarray}
satisfies equations \ (\ref{E1}) and (\ref{E2}), but for the case
of a zero  cosmological constant.

The above remarks \ suggest to check whether a composite
configuration, coinciding with the above described solutions in
the internal and the
external regions, globally satisfies the Einstein equations \ (%
\ref{E1}) and (\ref{E2}) for a cosmological term being constant
inside the mentioned sphere and vanishing out of it.

To examine this question, let us define the ansatz for
$u=1/g_{rr}$ and $v=-g_{00},$ for all values of the radial
distance by:
\begin{eqnarray}
u({r}) &=&\left(1-\frac{{\,{r}}_{0}}{r}\right)\text{ }\theta
(r-r_{0}){+}\left(1-\frac{\phi
_{0}^{2}\,r^{2}}{6}\right)\text{ }\theta (r_{0}-r){,} \nonumber \\
v({r}) &=&u({r}),\label{u}
\end{eqnarray}
where the constraint of $\ $making \ $u$ \ and $v$ \ to vanish in
the limits taken  from both sides at $ r=r_{0}$ has been imposed.
This condition determines \ $ r_{0}$ \ in terms of the
cosmological constant through
\begin{equation}
\frac{\phi _{0}^{2}\,r_{0}^{2}}{6}=1.  \label{ro}
\end{equation}
Henceforth, the derivative of \ $u$ \ takes the explicit form
\begin{eqnarray}
u^{\prime }(r) &=&\left(\frac{{\,}r_{0}}{r^{2}}\right)\text{ }\theta (r-r_{0}){+}\left(1-%
\frac{{\,}r_{0}}{r}\right)\text{ }\delta (r-r_{0}){-} \nonumber \\
&&\frac{\phi _{0}^{2}\,r}{3}\theta (r_{0}-r){-}\left(1-\frac{{{\mathit{\phi _{0}}%
^{2}}}\,r^{2}}{6}\right)\text{ }\delta (r_{0}-r){,}  \nonumber \\
&=&\left(\frac{{\,}r_{0}}{r^{2}}\right)\text{ }\theta (r-r_{0})-\frac{{{\mathit{%
\phi _{0}}^{2}}}\,r}{3}\text{ }\theta (r_{0}-r){,}
\end{eqnarray}
where the terms of Dirac's delta function  cancel precisely owing
to the selected condition (\ref{ro}). \

After substituting $u$ and $u^{\prime }\ $\ in the Einstein
equation \ (\ref {E1}), it follows that
\begin{widetext}
\begin{eqnarray*}
\frac{u^{\prime }(r)}{r}-\frac{1-u(r)}{r^{2}}{+}\Lambda (r) &=&\left(-\frac{{{%
\mathit{\phi _{0}}^{2}}\,}}{3}\right)\theta (r_{0}-r){+}\left(\frac{\,r_{0}}{r^{3}}%
\right)\theta (r-r_{0}){+} \\
&&\frac{1}{r^{2}}\left(1-\frac{{{\mathit{\phi
_{0}}^{2}}\,}r^{2}}{6}\right)\theta
(r_{0}-r)+\frac{1}{r^{2}}\left(1-\frac{{\,}r_{0}}{r}\right)\theta (r-r_{0}){+} \\
&&{({\frac{{{\mathit{\phi _{0}}^{2}}\,}}{2})}}\theta (r_{0}-r){-}\frac{1}{%
r^{2}}, \\
&\equiv &0.
\end{eqnarray*}
\end{widetext} \ \ \

\begin{figure*}
\includegraphics{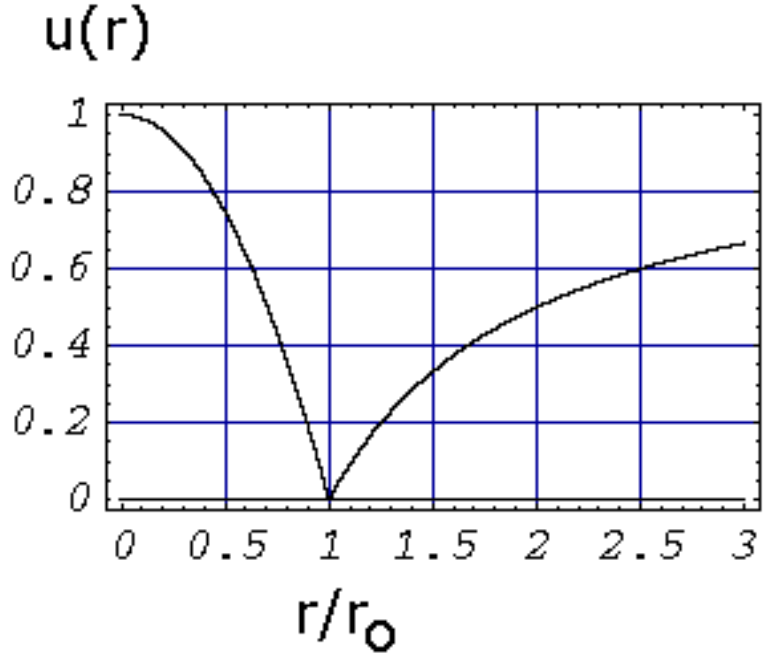}% Here is how to import EPS art
\caption{\label{fig1cosmo}   The common radial dependence of the
functions $u(r)$ and $v(r)$. Note the abrupt change in the slope
that is produced by  the sudden change in the cosmological
constant.}
\end{figure*}

Therefore, an exact solution of the couple of \ Einstein equations (\ref{E1}%
) and (\ref{E2}), in the sense of the distribution functions can
be written, in the simple form: \
\begin{eqnarray}
u(r) &=&v(r)=\left(1-\frac{{\,}r_{0}}{r}\right)\theta (r-r_{0}){+}\left(1-\frac{{\,}r^{2}}{%
r_{0}^{2}}\right)\theta (r_{0}-r){,}  \label{uv} \nonumber \\
r_{0} &=&{}\frac{\sqrt{6}}{|{{\mathit{\phi _{0}}}}|}.
\end{eqnarray}
Fig.\ref{fig1cosmo} illustrates the radial dependence of the
$u(r)=v(r)$ functions. It becomes clear that there is, for
example, a finite change in the slope of the $g_{00}$ component of
the metric. Thus a singularity is a associated to the boundary.

\ Concerning  the resting equation (\ref{E3}), it is not clear
that it can be satisfied.  Its non-linear nature is the main
source of the difficulty, since the candidate solutions (\ref{uv})
for $\ u$ and $v$ \ and
their derivatives are singular quantities,  and their products at the point \ $%
r_{0}$ are not well defined. In spite of this, the above
discussion leads to the expectation that a weak solution could
exists, showing the considered sudden change in the cosmological
term.

\section{Solubility criterion}

The results of the previous section suggest the existence of
solutions of the Einstein equations in the presence of a
cosmological constant which, suddenly reduces to zero outside a
sphere of radius \ $r_{0}.$

In this section we will argue that this system of equations can be
solved in a concrete sense to be defined below.

\bigskip
{\bf Definition}
\bigskip
{\it

 \noindent Consider the linear functionals in the space
$D^{\prime} \ $of test functions $f $   (see \cite{vladimirov})
and the generalized functions  $u_{g}$ and $v_{g}$ defined in
$D^{\prime} \ $ by a given succession $S$\ of fields
$(u_{n}(r),v_{n}(r)),n=1,2,3,...\infty $ through
\begin{eqnarray*}
u_{g} &=&\lim_{n\rightarrow \infty }\int_{0}^{\infty }dr\text{ }f(r)u_{n}(r),
\\
v_{g} &=&\lim_{n\rightarrow \infty }\int_{0}^{\infty }dr\text{ }f(r)v_{n}(r).
\end{eqnarray*}
We will say that the set of equations (\ref{E1})-(\ref{E3}) are
solved in a weak sense by ($u_{g}$,$v_{g})$, \ if the\ three
successions of\ \ linear \
functionals associated to the external sources $J_{0}^{0},J_{r}^{r}$ and $%
J_{\phi }^{\phi }=J_{\theta }^{\theta }$ \ and determined by each field
configuration $(u_{n}(r),v_{n}(r))$ through
\begin{eqnarray}
F_{0}^{0}[u_{n},v_{n}] &=&\int_{0}^{\infty }dr\text{ }%
f(r)J_{0}^{0}(u_{n},v_{n}),  \label{sources} \\
F_{r}^{r}[u_{n},v_{n}] &=&\int_{0}^{\infty }dr\text{ }%
f(r)J_{r}^{r}[u_{n},v_{n}],  \nonumber \\
F_{\phi }^{\phi }[u_{n},v_{n}] &=&F_{\theta }^{\theta }[u_{n},v_{n}],
\nonumber \\
&=&\int_{0}^{\infty }dr\text{ }f(r)J_{\phi }^{\phi }(u_{n},v_{n})\text{, \ }%
 \text{for all} \text{ }f\in D^{\prime },  \nonumber
\end{eqnarray}
all vanish in the limit $n \rightarrow \infty .$ }

It should be stressed that any field configuration can be
considered as\ a solution of the Einstein equations associated to
its corresponding external sources. These sources are nothing
other  than the result of evaluating the given fields in the
equations. \ Therefore the above definition declares as a solution
of the equations in the absence of sources, a generalized function
defined by a \ particular succession of field configurations,
whenever it turns out that the sources for the configurations
forming the  succession converge to zero in the sense of the
generalized functions. \ Such a definition
implies, in particular,  the physically desirable property that the limit $%
n\rightarrow \infty $ of the variation of the\ action defining the
equations, after being evaluated in the fields $(u_{n},v_{n})$
tends to \ vanish.

In the next section we will show that the given solubility
criterion for the equations (\ref{E1})-(\ref{E3}) \ is satisfied\
\ by a specially chosen succession of functions
$(u_{n}(r),v_{n}(r)),$ $\ n=1,2,3,...,\infty .$

For the purposes of the next section, we will consider a useful
representation of any spherically symmetric field configuration of
the Einstein equations. It can be found for example in
 Ref. \cite{synge}, and will be employed here as specialized for the
static situation under consideration. \ The first two equations
(\ref{E1}) and (\ref{E2}) allow us to obtain, for the functions
$\lambda $ $\ $and $\nu $ in (\ref{nula}):
\begin{eqnarray}
\lambda (r) &=&-\log {\large (}1-\frac{1}{2}\int_{0}^{r}r^{2}G_{0}^{0}(r)%
\text{ }dr{\large )},  \label{nula1} \\
\nu (r) &=&\int_{0}^{r} dr {\Huge(}\frac{\exp (\lambda
(r))-1}{r}-r\exp (\lambda (r))G_{r}^{r}(r){\Huge)}, \nonumber \\
\label{expnu}
\end{eqnarray}
expressing these functions in terms of the two components of the
Einstein tensor $G_{r}^{r}(r)$ and $G_{0}^{0}(r).$ Then, the field
configurations $(u,v)$ are also fully defined as functions of
$G_{1}^{1}(r)$ and $G_{0}^{0}(r)$ by  (\ref{nula}) in the form
\begin{eqnarray}
v(r) &=&g_{00}(r)=\exp (\nu (r)),  \label{uv1} \\
u(r) &=&-g_{rr}(r)=\exp (-\lambda (r)).  \label{uv2}
\end{eqnarray}

\ Moreover, the vanishing of the covariant divergence of the Einstein tensor
as an identity, allows the expression of  the components $G_{\phi }^{\phi }$ a $%
G_{\theta }^{\theta }$ \ as functions of  $G_{1}^{1}(r)$ and $%
G_{0}^{0}(r)$ through \cite{synge}
\begin{equation}
G_{\phi }^{\phi }=G_{\theta }^{\theta }=\frac{r}{2}\partial _{r}G_{r\text{ }%
}^{r}(r)+G_{r}^{r}(r)+\frac{r}{4}\partial _{r}\nu
(r)(G_{r}^{r}(r)-G_{0}^{0}(r)).\   \label{angular}
\end{equation}

As discussed in Ref. \cite{synge}, the meaning of the written
expressions is that it becomes possible to determine the metric by
arbitrarily\ fixing the two components $G_{1}^{1}(r)$ and
$G_{0}^{0}(r)$ of the Einstein tensor, and afterwards simply
select the resting components $G_{\phi }^{\phi }$ and $G_{\theta
}^{\theta }$ as defined by (\ref{angular}) in order to complete
the satisfaction of the Einstein equations. Clearly, in these
equations the energy-momentum tensor is assumed to be equal to the
Einstein tensor, fixed in the explained way.

 \ Let us now consider that all three functions $\Lambda $ are
equal to a common one, $\Lambda (r)$,  for all the radial axis
except within a small open neighbourhood $B$ of the boundary point
$r_{0}.$ Then,   the eq. (\ref{angular}) implies, for these
regions:
\[
\partial _{r}\Lambda (r)=0.
\]

Thus, the function $\Lambda (r)$ should be strictly constant in
each of the two zones. \ However, the possibility is not yet
discarded that the two\ values associated to each of the two
disjoint regions in which $r_{0}$ divides the radial axis could be
different. Let us look into this. \ In the case that they could
effectively differ, the \ term $\partial _{r}G_{r\text{
}}^{r}(r)=\partial _{r}\Lambda (r)$ in eq. (\ref{angular}) will
contribute with a Dirac delta like singularity. Thus, whenever the
factor $\partial _{r}\nu (r)$ is a regular one in the
neighbourhood  of $r_{0},$ there will be no
possibility other than the coincidence of the two values of the function $%
\Lambda .$ \ However, since the factor $\partial _{r}\nu (r)$ can
show singularities  at some special points,  the opportunity is
yet open for cancelling the Dirac delta term. This is only
possible by breaking the equality between $\Lambda _{r\text{ }}$
and $\Lambda _{0}$, which allows the mentioned singular dependence
of the factor $\partial _{r}\nu (r)$  to play a role near special
points. \ \ In this way the Einstein equations could be obeyed in
the sense of the generalized functions.

In the next section the satisfaction of \ the Einstein equations
(\ref {E1})-(\ref{E3})  will be discussed, in accordance with the
given solubility criterion, when all the $\Lambda $ \ functions
have a Heaviside step function \ behavior. \bigskip

\section{The solution}

Let us search  in this section  a  solution of the equations (\ref
{E1})-(\ref{E3})  for the specific form of the cosmological term
suggested by the discussion in Section 2, that is
\begin{eqnarray}
\Lambda _{r}(r) &=&\Lambda _{r}(r)=\Lambda (r), \label{cosmo} \\
&=&\Lambda \text{ }\theta (r_{0}-r).  \nonumber
\end{eqnarray}

The first step in showing the satisfaction of the  solubility
criterion by a
succession of functions $(u_{n},v_{n})\ $tending to (\ref{u}) in the limit $%
n\rightarrow \infty $ \,  will be to  define a corresponding
succession of the values for the radial \ and temporal components
of the Einstein tensor. These components will be selected to
approach the step-like cosmological function (\ref{cosmo}) in the
limit $n\rightarrow \infty $ as:
\begin{eqnarray}
G_{r}^{r}(r|n) &=&\Lambda \text{ }\theta _{n}^{(1)}(r_{0},r),
\label{g11n}
\\
G_{0}^{0}(r|n) &=&\Lambda \text{ }\theta _{n}^{(0)}(r_{0}-r),  \label{g00n}
\\
\sigma_{n}(r) &=& \theta _{n}^{(1)}(r_{0}-r){-}\theta
_{n}^{(0)}(r_{0}-r),
\end{eqnarray}
where the succession of functions $\theta _{n}^{(1)}$ and $\theta
_{n}^{(0)}$ for all values of $\ n$ \ are both chosen to define
the step functions appearing in (\ref{cosmo}) in  the limit $n
\rightarrow \infty $. \ The precise expression for $\theta
_{n}^{(0)}$ will be
\[
\text{ }\theta _{n}^{(0)}(r_0-r)=\theta (r_{0}-\epsilon ^{\prime
}(n)-r),
\]
where $\theta (x)$ is the Heaviside step function and \ $\epsilon
^{\prime }(n)$ is a small quantity with respect to $r_{0}$, which
is taken as vanishing in the limit $n\rightarrow \infty .$ \ The
regularization for $\theta _{n}^{(1)}(r_{0},r)$ is chosen as given
by
\begin{eqnarray}
\theta _{n}^{(1)}(r_{0},r) &=&\theta \text{(}r_{0}-\epsilon
^{\prime }(n)-r)+\sigma
_{n}(r),  \label{sigma} \\
\sigma _{n}(r) &=&g_{n}(r)\theta (r-r_{0}+\epsilon ^{\prime }(n))\theta
(r_{0}-\epsilon (n)-r),  \nonumber \\
\epsilon ^{\prime }(n) &>&\epsilon (n),  \nonumber
\end{eqnarray}
where $\epsilon (n)$ is another  segment, also tending to zero in
the limit $n\rightarrow \infty ,$ but  smaller as \ a real number than $%
\epsilon ^{\prime }(n).$ The up to now arbitrary (but assumed to be bounded)
function $g_{n}(r))$ will be fixed in what follows.

Therefore, for this $n$-dependent selection of the two arbitrary
components of the Einstein tensor, $G_{0}^{0}$ and $G_{r}^{r}$,
the quantities $u_{n}$ and $v_{n}$ determined through using
(\ref{nula}) and (\ref{nula1}), are solutions of the
Einstein equations whenever the angular components are calculated from (%
\ref{angular}).  Thus, in order to show that the succession
$(u_{n},v_{n})$ \ satisfies the solubility criterion for the
space-dependent cosmological constant (\ref{cosmo}), it is only
needed to prove that the generalized functions associated to their
external sources (\ref{sources}) have vanishing limits $%
n\rightarrow \infty $. Let us show  this property below. \

The linear functionals in $D^{\prime} $ being equivalent  to the
auxiliary external sources for which each of the pairs
$(u_{n},v_{n})$ turns to be a solution of \ Einstein equations
(with cosmological term defined by (\ref{cosmo}))  can be
explicitly written as
\begin{eqnarray}
F_{0}^{0}[u_{n},v_{n}] &=&\int_{0}^{\infty }dr\text{ }f(r)\text{ }%
J_{0}^{0}(u_{n},v_{n}),  \nonumber \\
&=&\int_{0}^{\infty }dr\text{ }f(r)\left( -\Lambda
(r)+G_{0}^{0}(r|n)\right)
,  \nonumber \\
&=&\int_{0}^{\infty }dr\text{ }f(r)\left( -\Lambda
(r)-\frac{{u_{n}^{\prime
}(}r{)}}{r}+\frac{1-u_{n}(r)}{r^{2}}\right) ,  \label{f00} \\
F_{r}^{r}[u_{n},v_{n}] &=&\int_{0}^{\infty }dr\text{ }%
f(r)J_{r}^{r}(u_{n},v_{n}),  \nonumber \\
&=&\int_{0}^{\infty }dr\text{ }f(r)\left( -\Lambda
(r)+G_{r}^{r}(r|n)\right)
,  \nonumber \\
&=&\int_{0}^{\infty }dr\text{ }f(r)\left( -\Lambda (r)-\frac{u_{n}(r)}{%
v_{n}(r)}\,\frac{{v_{n}^{\prime
}(r)}}{r}+\frac{1-u_{n}(r)}{r^{2}}\right) ,
\label{frr} \\
F_{\phi }^{\phi }[u_{n},v_{n}] &=&\int_{0}^{\infty }dr\text{
}f(r)J_{\phi }^{\phi }(u_{n},v_{n})=\int_{0}^{\infty }dr\text{
}f(r)J_{\theta }^{\theta
}(u_{n},v_{n}),  \nonumber \\
&=&\int_{0}^{\infty }dr\text{ }f(r)(-\Lambda (r)+G_{\phi }^{\phi
}(r|n)),
\nonumber \\
&=&\int_{0}^{\infty }dr\text{ }f(r) \left( -\Lambda (r)-\frac{u_{n}(r)}{{2}%
} \left( \frac{v_{n}^{^{\prime \prime
}}(r)}{v_{n}(r)}-\frac{v_{n}^{\prime }(r)^{2}}{2v_{n}(r)^{2}}+
\frac{v_{n}^{\prime }(r)\text{ }u_{n}^{\prime }(r)}{2\text{
}v_{n}(r)\text{
}u_{n}(r)}+\frac{{1}}{r}(\frac{u_{n}^{\prime }(r)}{u_{n}(r)}+\frac{%
v_{n}^{\prime }(r)}{v_{n}(r)}) \right) \right),  \label{f33} \\
\Lambda (r) &=&\Lambda \text{ }\theta (ro-r).
\end{eqnarray}

Our purpose in what follows will be to show that these functionals
tend to zero by properly selecting the regularization, that is, to
argue that there exists a succession of field configurations,
approaching the fields (\ref{uv}%
)\ in the sense of the generalized functions, for which the
Einstein tensor also tends to the step-like cosmological term \
(\ref{cosmo}) in the same sense. \

\ For eq.(\ref{f00}) and  (\ref{frr}), thanks to the same
definitions of the succession of Einstein tensors
(\ref{g11n}),(\ref{g00n}), it follows that
\begin{widetext}
\begin{eqnarray}
\lim_{n\rightarrow \infty }F_{0}^{0}[u_{n},v_{n}] &=&\lim_{n\rightarrow
\infty }\int_{0}^{\infty }dr\text{ }f(r)\left( -\Lambda
(r)+G_{0}^{0}(r|n)\right) ,  \nonumber \\
&=&\lim_{n\rightarrow \infty }\int_{0}^{\infty }dr\text{
}f(r)\left( -\Lambda (r)+\Lambda \text{ }\theta (r_{0}-\epsilon
^{\prime }(n)-r) \right) ,
\nonumber \\
&=&-\Lambda \lim_{n\rightarrow \infty }\int_{r_{0}-\epsilon ^{\prime
}(n)}^{r_{0}}dr\text{ }f(r),  \nonumber \\
&=&0, \\
\lim_{n\rightarrow \infty }F_{r}^{r}[u_{n},v_{n}] &=&\lim_{n\rightarrow
\infty }\int_{0}^{\infty }dr\text{ }f(r)\left( -\Lambda
(r)+G_{r}^{r}(r|n)\right) ,  \nonumber \\
&=&\lim_{n\rightarrow \infty }\int_{0}^{\infty }dr\text{ }f(r)\left(
-\Lambda (r)+\Lambda \text{ }\theta (r_{0}-\epsilon ^{\prime }(n)-r)+\sigma
_{n}(r)\right) ,  \nonumber \\
&=&-\Lambda \lim_{n\rightarrow \infty }\int_{r_{0}-\epsilon ^{\prime
}(n)}^{r_{0}}dr\text{ }f(r)+\lim_{n\rightarrow \infty }\int_{r_{0}-\epsilon
^{\prime }(n)}^{r_{0}-\epsilon (n)}dr\text{ }f(r)\text{ }\sigma _{n}(r),
\nonumber \\
&=&0,
\end{eqnarray}
\end{widetext}
where the last equality follows thanks to the fact that $\sigma
_{n}(r)$ is assumed to be a bounded function for all $\ n.$ \ \
Therefore, the first two
functionals associated to the external sources vanish in the limit $%
n\rightarrow \infty $. \ \ \

For the last functional it is possible to write first
\begin{widetext}
\begin{eqnarray}
\lim_{n\rightarrow \infty }F_{\phi }^{\phi }[u_{n},v_{n}]
&=&\lim_{n\rightarrow \infty }\int_{0}^{\infty }dr\text{ }f(r)(-\Lambda
(r)+G_{\phi }^{\phi }(r|n)),  \nonumber \\
&=&\lim_{n\rightarrow \infty }\int_{0}^{\infty }dr\text{ }f(r)(-\Lambda (r)+%
\frac{r}{2}\partial _{r}G_{r\text{ }}^{r}(r|n)+G_{r}^{r}(r|n)+\frac{r}{4}%
\partial _{r}\nu _{n}(r)(G_{r}^{r}(r|n)-G_{0}^{0}(r|n))\ ),  \nonumber \\
&=&\lim_{n\rightarrow \infty }\int_{0}^{\infty }dr\text{ }f(r)(\frac{r}{2}%
\partial _{r}G_{r}^{r}(r|n)+\frac{r}{4}\partial _{r}\nu
_{n}(r)(G_{r}^{r}(r|n)-G_{0}^{0}(r|n))\ ,  \nonumber \\
&=&\lim_{n\rightarrow \infty }\int_{0}^{\infty }dr\text{ }f(r)\left( \frac{r%
}{2}\partial _{r}{\large (}\Lambda \text{ }\theta (r_{0}-\epsilon
^{\prime }(n)-r)+\Lambda \text{ } \sigma _{n}(r){\large
)}+\frac{r}{4}\partial _{r}\nu
_{n}(r)(G_{r}^{r}(r|n)-G_{0}^{0}(r|n))\right) ,\   \nonumber \\
&=&-\frac{1}{2}\Lambda \text{ }f(r_{0})+\lim_{n\rightarrow \infty
}\int_{0}^{\infty }dr\text{ }f(r)\frac{r}{4}\partial _{r}\nu
_{n}(r)(G_{r}^{r}(r|n)-G_{0}^{0}(r|n)).  \label{f3}
\end{eqnarray}

\ \ But, from the general relations (\ref{nula1})-(\ref{uv2}) it
follows that
\begin{eqnarray}
\lambda _{n}(r) &=&-\log {\large (}1-\frac{1}{2}%
\int_{0}^{r}r^{2}G_{0}^{0}(r|n)\text{ }dr{\large )}, \\
\nu _{n}(r) &=&\int_{0}^{r} dr \left( \frac{\exp (\lambda _{n}(r))-1}{r}%
-r\exp \left( \lambda (r) \right) G_{1}^{1}(r|n) \right),
\end{eqnarray}
which, after taking the derivative of $\nu _{n}(r)$, gives
\begin{eqnarray*}
\partial _{r}\nu _{n}(r) &=&\frac{\exp ( \lambda _{n}(r))-1}{r}-r\exp
(\lambda _{n}(r))\, G_{1}^{1}(r|n), \\
&=&-\frac{1}{r}+\frac{(\frac{1}{r}-\Lambda r\theta _{n}^{(0)}(r_{0}-r))}{%
1-\int_{0}^{r}drr^{2}\theta _{n}^{(1)}(r_{0}-r)}.
\end{eqnarray*}

Further, the denominator in the last expression can be explicitly
evaluated as
\begin{eqnarray}
\exp (-\lambda _{n}(r)) &=&1-\int_{0}^{r}drr^{2}\theta
_{n}^{(1)}(r_{0}-r),
\label{deno} \\
&=&\theta (r_{0}-\epsilon ^{\prime }(n)-r) \left( 1-\frac{\Lambda
r^{2}}{3} \right)+\theta (r-r_{0}+\epsilon ^{\prime }(n))
\left(1-\frac{\Lambda }{r}\frac{(r_{0}-\epsilon ^{\prime
}(n))^{3}}{3} \right).  \nonumber
\end{eqnarray}
After inserting (\ref{sigma}) and (\ref{deno}) in (\ref{f3}) the
functional $F_{\phi}^{\phi}$ can be transformed in the following
way:
\begin{eqnarray*}
\lim_{n\rightarrow \infty }F_{\phi }^{\phi }[u_{n},v_{n}] &=&-\frac{1}{2}%
\Lambda \text{ }r_{0}\text{ }f(r_{0})+\lim_{n\rightarrow \infty
}\int_{0}^{\infty }dr\text{ }f(r)\frac{r}{4}\partial _{r}\nu
_{n}(r)(G_{r}^{r}(r|n)-G_{0}^{0}(r|n)), \\
&=&-\frac{1}{2}\Lambda \text{ }r_{0}\text{
}f(r_{0})-\lim_{n\rightarrow \infty }\int_{0}^{\infty }dr\text{
}f(r)\frac{\Lambda }{4} \sigma _{n}(r)+ \\
&&\frac{\Lambda }{4}\lim_{n\rightarrow \infty }\int_{r-r_{0}+\epsilon
^{\prime }(n))}^{r_{0}-\epsilon (n)-r}dr\text{ }f(r)\frac{\sigma
_{n}(r)(1-\Lambda r^{2}\sigma _{n}(r))}{(1-\frac{\Lambda }{r}\frac{%
(r_{0}-\epsilon ^{\prime }(n))^{3}}{3})}, \\
&=&-\frac{1}{2}\Lambda \text{ }r_{0}\text{ }f(r_{0})+\frac{\Lambda }{4}%
\lim_{n\rightarrow \infty }\int_{r-r_{0}+\epsilon ^{\prime
}(n))}^{r_{0}-\epsilon (n)-r}dr\text{ }f(r)\frac{\sigma
_{n}(r)(1-\Lambda r^{2}\sigma _{n}(r))}{(1-\frac{\Lambda
}{r}\frac{(r_{0}-\epsilon ^{\prime }(n))^{3}}{3})}.
\end{eqnarray*}

In order to proceed, we complete the specification of the forms of
$u_n$ and $v_n$ by  fixing  the functions $\sigma _{n}$ as given
by
\[
\sigma _{n}(r)=\frac{1}{6}\text{ }\theta (r_{0}-\epsilon (n)-r)\text{ }%
\theta (r-r_{0}+\epsilon ^{\prime }(n));
\]
after defining the new integration variable\ $\ z$ \ and
parameter\ $\Delta$ according to
\begin{eqnarray*}
z &=&\frac{1}{\epsilon ^{\prime }(n)}(r-r_{0}), \\
\epsilon (n) &=&-\Delta \text{ }\epsilon ^{\prime }(n),
\end{eqnarray*}
this allows us to write

\begin{eqnarray*}
\lim_{n\rightarrow \infty }F_{\phi }^{\phi }[u_{n},v_{n}] &=&-\frac{1}{2}%
\Lambda r_{0}f(r_{0})+ \\
&&\frac{\Lambda }{24}\lim_{n\rightarrow \infty }\int_{0}^{\infty }dz\text{ }%
f(r_{0}+\epsilon ^{\prime }(n)z)\frac{\frac{\Lambda }{3}(\frac{1}{2}%
r_{0}^{2}-\epsilon ^{\prime }(n)r_{0}z-\frac{1}{2}\epsilon ^{\prime
}(n)^{2}z^{2})(r_{0}+\epsilon ^{\prime }(n)\text{ }z)\theta (1+z)\theta (z-\Delta)%
}{\frac{\Lambda }{3}(z\text{ }r_{0}^{2}+3r_{0}^{2}-\epsilon ^{\prime
}(n)r_{0}-\epsilon ^{\prime }(n)^{2})}, \\
&=&-\frac{1}{2}\Lambda r_{0}f(r_{0})+f(r_{0})\frac{\Lambda \text{ }r_{0}}{48}%
\lim_{n\rightarrow \infty }\int_{-1}^{\Delta}dz\frac{1}{z+3}, \\
&=&f(r_{0})\frac{\Lambda r_{0}}{2}(-1+\frac{1}{24}\log
\frac{(\Delta+3)}{2}).
\end{eqnarray*}

In this way, after selecting $\Delta$ defined by
\[
\Delta=2\exp (24)-3,
\]
it follows that the succession of regularized fields $\
S=\{(u_{n},v_{n})\}$ \ has a corresponding succession of
associated externals sources, which  vanish  in the limit
$n\rightarrow \infty. $ Therefore the generalized function
$(u_{g},v_{g})$ defined with the precision of being the limit of
the linear functionals associated to the specific configurations
in $S$,  satisfies the Einstein equations in the  weak sense
defined here. It can noticed that for linear systems this
definition is less restrictive and the class of successions of
field configurations allowed for expressing $(u_{g},v_{g})$ as
their limit is very much more wider.

\section{\protect\smallskip Conclusions}

\ A criterion for a generalized function to be a solution of the
non-linear Einstein equations is proposed. Then,  a particular
solution satisfying the criterion for the Einstein equations in
the presence of a cosmological term which suddenly vanishes
outside a given radial distance, is found. The considered
space-time shows a homogeneous deSitter Universe being at an
internal region and the Schwartzschild space for the external one.
The characterization of the appearing  singularity at the boundary
will be considered in future works.

\bigskip

\section*{Acknowledgements}

The helpful conversations and remarks of R. F. Parada, F. Ongay,
O.P. Fern\'andez, R. Acharya, L. Jiayu and L. Dung Trang  during
the preparation of the work are gratefully acknowledged.  The
careful revision of Ms. Suzy Vascotto at the TH Division of CERN
Secretariat is also very much appreciated.

\bigskip

\bigskip
\end{widetext}

\end{document}